# Hierarchical Optimal Dispatch of Active Distribution Networks Considering Flexibility Auxiliary Service of Multi-community Integrated Energy Systems

Chunling Wang, Chunming Liu, Xiulin Zhou, Yang Li, Senior *Member, IEEE,* and Gaoyuan Zhang

*Abstract*—Active distribution networks (ADNs) are the main platforms for carrying large-scale distributed renewable energy and flexible resources, and multi-community integrated energy systems (MCIESs) may become important flexible resource supplies in ADNs owing to their multi-energy synergistic and complementary advantages. To fully utilize the flexible regulation potential of MCIESs for ADNs, a novel hierarchical stochastic dispatch approach for ADNs that considers flexibility auxiliary services of MCIESs is proposed. In this approach, a flexibility auxiliary service pricing strategy that combines adjustment cost and flexibility margin is established by evaluating the operational flexibility of MCIESs. In addition, considering renewable uncertainty, an MCIES-ADN flexibility interaction mechanism based on insufficient flexibility risk is designed to optimize their operation strategies and reduce the uncertainty risk. In the solution phase, an analytical target cascading theory-based distributed solving method is developed to realize decoupling and parallel solving of multiple stakeholders. The simulation results for a PG&E 69-node system with three CIESs demonstrate that the proposed approach not only improves MCIES revenue but also enhances ADN flexibility to consume renewable energy, which provides a fundamental way for efficient application of regional mutual aid.

## I. INTRODUCTION

WITH the vigorous promotion of energy conservation and emissions reduction, active distribution networks (ADNs) that can carry large-scale distributed renewable energy generators (REGs) have been rapidly developed [1]. However, owing to the strong volatility and uncertainty of REG outputs, ADNs require sufficiently flexible resources to ensure efficient consumption of renewable energy and reliable power supply [2]. Simultaneously, an increasing number of community integrated energy systems (CIESs) are being integrated into ADNs due to their multi-energy complementarity and ladder utilization abilities [3]. For ADNs, multi-community integrated energy systems (MCIESs) have bidirectional regulation effects on the electric power, and their internal plentiful controllable devices can serve as flexible resources to cope with REG volatility through power interactions, thereby improving the flexibility of ADNs. In addition, an ADN with MCIESs contains multiple stakeholders and numerous types of distributed power generators [4], making it challenging to break from the existing mode of independent operation of each energy supply system to realize integrated optimized operation while considering the interests of all stakeholders. Therefore, from the perspective of flexible dispatch of ADNs, exploring how to fully utilize the flexible support potential of MCIES for ADNs and balance the interests of multiple stakeholders is of great significance for promoting large-scale connections and efficient utilization of renewable energy.

*A. Literature Review*

Given the trend toward a high proportion of distributed REGs, considerable research has explored the optimal dispatch of ADNs to improve flexibility. Reference [5] utilized soft open points and battery energy storage (BES) to improve the operational flexibility of an ADN. In [6], photovoltaics (PV), BES, and the flexibility market were jointly optimized to provide flexibility. Reference [7] exploited flexibility from the network–storage–load demand response (DR) in an ADN to enhance the operating cost, voltage distribution, and risk control. The operation of distributed generators was optimized in [8] to improve the distribution network flexibility. The above studies effectively enhanced the operational flexibility of ADNs by utilizing their internal flexible resources. However, in the current context of vigorously developing multi-energy coupling and flexible interconnection technologies, few studies have considered the flexibility support ability of connected MCIESs which have plentiful controllable devices in ADN optimal scheduling research.

Additionally, the uncertainty of REG outputs is considered a key issue in system optimal dispatch, and currently statistical methods and machine learning are commonly used to model the uncertainty. For example, reference [9] proposed probabilistic models for wind turbine (WT) and PV with the discretization of random variables. Copula theory was utilized in [10] to establish a joint probabilistic output distribution model for WT and PV. Reference [11] built a generative adversarial network to learn the characteristics of REG outputs. Reference [12] proposed a data-driven Bayesian nonparametric approach to construct REG uncertainty sets. Although these studies effectively modeled REG uncertainty during system

Chunling Wang, Chunming Liu and Xiulin Zhou are with the School of Electrical and Electronic Engineering, North China Electric Power University, Beijing 102206, China, (e-mail: wangcl@ncepu.edu.cn; liuchunming@ncepu.edu.cn; zhouxiulin2022@163.com).
Yang Li is with the School of Electrical Engineering, Northeast Electric Power University, Jilin 132012, China, (e-mail: liyang@neepu.edu.cn).
Gaoyuan Zhang is with the Key Laboratory of Power Station Energy Transfer Conversion and System (Ministry of Education), North China Electric Power University, Beijing 102206, China, (email: gaoyuan@ncepu.edu.cn).

optimization, they disregarded the risk of insufficient flexibility in uncertain environments, which may lead to sub-optimality.

In terms of model construction and solving, the dynamic dispatch problem of ADN containing MCIESs can be described by a nonlinear optimization model, which is divided into centralized and distributed modeling approaches. Centralized modeling is to associate the relevant constraints of ADN and MCIES into the overall system model, which can be solved holistically using mixed integer programming [13], sequential quadratic programming [14], and neural networks [15]. However, the above methods require the collection of each device parameter and internal information of MCIES, resulting in a large amount of transmitted information and difficulty in reflecting the different interests of MCIES and ADN. The distributed modeling method models ADN and each CIES as different stakeholders, which is currently a development trend in this field. For example, reference [16] established a Stackelberg game-based optimal scheduling model for MCIES and solved it using particle swarm algorithm. A distributed iterative solution method based on a metaheuristic was designed in [11] to solve the MCIES interaction problem.

The aforementioned studies produced excellent efforts in the optimal dispatch of ADNs. However, some research gaps still exist in this field: (1) MCIESs have plentiful and diversified flexible controllable resources, although their flexibility regulation potential for ADNs has not been fully explored; (2) The risk of insufficient flexibility is aggravated under a high percentage of REGs, however, previous studies focused on portraying REG uncertainties but neglected to perform in-depth analyses or quantification of the potential risks they pose to the system; (3) For complex systems containing multiple stakeholders, centralized modeling methods have simple and clear ideas, but cannot protect the privacy of each stakeholder and reflect their decentralized autonomy characteristic.

*B. Contributions*

In response to the above issues, this study proposes a hierarchical optimal dispatch approach for ADN focusing on the flexibility auxiliary service (FAS) of MCIESs to fully exploit the flexibility benefits of connected systems under uncertain environments. The main contributions are as follows:
1) To fully utilize the flexibility regulation potential of MCIESs for ADNs, a novel hierarchical stochastic optimal dispatch model of ADN considering FAS of MCIESs is proposed, in which the flexibility margin of MCIES is effectively evaluated and an FAS pricing strategy that incorporates the regulation cost and flexibility margin is designed.
2) An ADN-MCIES interaction mechanism considering the risk of insufficient flexibility is developed to optimize their operation strategies, which manages to reduce the uncertainty risk of REG outputs and optimize flexibility interactions among multiple stakeholders.
3) A distributed iterative solution method based on the analytical target cascading (ATC) theory is presented to decouple and efficiently solve MCIESs in parallel while only transmitting necessary data, which effectively protects the privacies of all parties and ensures convergence of the solution.

## II. Flexibility Evaluation of CIES

The purpose of flexibility evaluation is to grasp the flexibility margins of CIES dispatch in meeting its own energy demand at each time period, so as to provide flexible resources and formulate FAS prices for transactions with ADN. Therefore, a two-stage CIES flexibility evaluation is first performed, involving pre-dispatch under an "isolated network" followed by flexibility margin modeling. Fig. 1 presents the overall structure and energy flow of a typical CIES analyzed in this study.

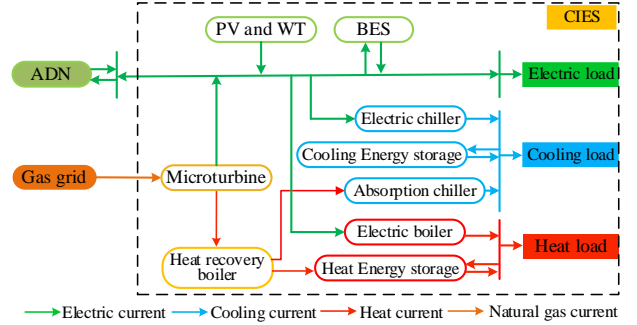

**Fig. 1.** Overall structure and energy flow of a typical CIES

*A. Stage 1: Pre-dispatch of CIES*

The objective of CIES pre-dispatch is to minimize the total operating cost, including fuel cost of microturbine (MT) $C^1_{CIES}$, operation and maintenance (O&M) cost $C^2_{CIES}$, DR compensation cost $C^3_{CIES}$, renewable curtailment and load shedding penalty cost $C^4_{CIES}$.

$$\min C^{pre}_{CIES} = C^1_{CIES} + C^2_{CIES} + C^3_{CIES} + C^4_{CIES}$$
$$C^1_{CIES} = \sum_{t=1}^{T} \gamma_{gas}(P_{mt,t}/\eta_{mt}/H_{gas})\Delta t$$
$$C^2_{CIES} = \sum_{t=1}^{T}\sum_{i\in\Lambda_d} P_{i,t}\kappa_i\Delta t \qquad (1)$$
$$C^3_{CIES} = \sum_{t=1}^{T}[\gamma_{DR}\times\min(P_{tse,t},0)]\Delta t$$
$$C^4_{CIES} = \sum_{t=1}^{T}(\gamma_{cr}P_{cr,t}+\gamma_{cl}P_{cl,t})\Delta t$$

where $\gamma_{gas}$ is the natural gas price; $P_{mt,t}$ is the electric output of MT during period $t$; $\eta_{mt}$ and $H_{gas}$ are the electricity conversion factor and calorific value of MT, respectively; $\Lambda_d$ is the set of operating devices; $P_{i,t}$ and $\kappa_i$ are the output power and O&M cost factor of device $i$ during period $t$; $\gamma_{DR}$ is the DR compensation cost factor; $P_{tse,t}$ is the time-shiftable electric (TSE) load during period $t$; $P_{cr,t}$ and $P_{cl,t}$ are the renewable curtailment and load shedding power during period $t$, $\gamma_{cr}$ and $\gamma_{cl}$ are the penalty cost factors; $\Delta t$ is the dispatch interval, and $T$ is the dispatch cycle.

The operational constraints in the pre-dispatch stage include:

*1) Energy-producing Device Constraints*

The controllable generator that can actively supply energy in CIESs is called energy-producing device, i.e., MT. It should satisfy the following constraints [17].

$$\begin{cases} P_{mt}^{\min}\delta_{mt,t} \leq P_{mt,t} \leq P_{mt}^{\max}\delta_{mt,t} \\ -r_{mt}^d \leq P_{mt,t} - P_{mt,t-1} \leq r_{mt}^u \\ \delta_{mt,t} - \delta_{mt,t-1} \leq \delta_{mt,\tau_1}, \quad \tau_1 = t+1,\ldots,\min(t+T_{mt}^{up}-1,T) \\ \delta_{mt,t-1} - \delta_{mt,t} \leq 1 - \delta_{mt,\tau_2}, \quad \tau_2 = t+1,\ldots,\min(t+T_{mt}^{down}-1,T) \end{cases} \quad (2)$$

where $P_{mt}^{\max}$ and $P_{mt}^{\min}$ are the maximum and minimum electric outputs of MT, respectively; $\delta_{mt,t}$ is either 0 (shutdown state) or 1 (startup state); $r_{mt}^d$ and $r_{mt}^u$ are the downward and upward ramp rates; $\tau_1$ and $\tau_2$ are auxiliary variables constrained by the minimum operating time $T_{mt}^{up}$ and stopping time $T_{mt}^{down}$ of MT.

*2) Coupling Device Constraints*

Controllable units that can realize various types of energy conversion in CIESs are called coupling devices and include heat recovery boiler (HRB), electric chiller (EC), absorption chiller (AC) and electric boiler (EB). The coupling device constraints are expressed as follows [18]:

$$\begin{cases} P_i^{\min} \leq P_{i,t}^{in} \leq P_i^{\max} \\ -r_i^d \leq P_{i,t}^{in} - P_{i,t-1}^{in} \leq r_i^u \\ P_{i,t}^{out} = \eta_i P_{i,t}^{in} \end{cases} \quad (3)$$

where $P_{i,t}^{in}$ and $P_{i,t}^{out}$ are the generalized expressions for the input and output powers of coupling device $i$; $r_i^d$ and $r_i^u$ are the downward and upward ramp rates, respectively; $\eta_i$ is the conversion efficiency.

In particular, the input of HRB is the heat produced by MT, which obeys the following relationship:

$$H_{hrb,t}^{in} = \frac{P_{mt,t}(1-\eta_{mt}-\eta_{mt}^{loss})}{\eta_{mt}} \quad (4)$$

where $\eta_{mt}^{loss}$ is the heat loss rate of MT.

*3) Energy Storage Device Constraints*

Energy storage devices in CIESs should satisfy charge/discharge and capacity constraints, which are expressed as follows [19]:

$$\begin{cases} S_{i,t} = (1-k_{i,loss})S_{i,t-1} + (\eta_i^{ch} P_{i,t}^{ch} - P_{i,t}^{dc}/\eta_i^{dc})\Delta t \\ 0 \leq P_{i,t}^{dc} \leq P_i^{dc,\max} \\ 0 \leq P_{i,t}^{ch} \leq P_i^{ch,\max} \\ S_i^{\min} \leq S_{i,t} \leq S_i^{\max} \\ S_i^0 = S_i^{T_{end}} \end{cases} \quad (5)$$

where $S_{i,t}$ is the capacity of energy storage device $i$ during period $t$; $k_{i,loss}$ is the energy loss coefficient; $\eta_i^{ch}$ and $\eta_i^{dc}$ are the charge and discharge power efficiency; $P_i^{ch,\max}$ and $P_i^{dc,\max}$ are the rated charge and discharge power of energy storage device $i$; $S_i^{\min}$ and $S_i^{\max}$ are the minimum and maximum capacity; $S_i^0$ and $S_i^{T_{end}}$ are the initial and end-of-scheduling capacity, respectively.

*4) DR Constraints*

The TSE load needs to satisfy the following constraints:

$$\begin{cases} -\alpha_{tse} P_{l,t} \leq P_{tse,t} \leq \alpha_{tse} P_{l,t} \\ \sum_{t=1}^{T} P_{tse,t} = 0 \end{cases} \quad (6)$$

where $\alpha_{tse}$ is the ratio of the TSE load, and $P_{l,t}$ is the original electric load during period $t$.

*5) Power Balance Constraints*

The supply and demand balance of electric, heat, and cooling energies must be guaranteed to ensure the safe and stable operation of the CIES.

$$\begin{cases} P_{wt,t} + P_{pv,t} + P_{mt,t} + P_{bes,t}^{dc} = P_{l,t} + P_{tse,t} + P_{bes,t}^{ch} + P_{eb,t}^{in} + P_{ec,t}^{in} - P_{cl,t} \\ H_{hrb,t}^{out} + H_{eb,t}^{out} + H_{hes,t}^{dc} = H_{l,t} + H_{hes,t}^{ch} + H_{ac,t}^{in} \\ C_{ac,t}^{out} + C_{ec,t}^{out} + C_{ces,t}^{dc} = C_{l,t} + C_{ces,t}^{ch} \end{cases} \quad (7)$$

where $P_{wt,t}$ and $P_{pv,t}$ are the power consumed by the WT and PV during period $t$, respectively; $H_{l,t}$ and $C_{l,t}$ are the heat and cooling loads during period $t$, respectively.

*6) Renewable Consumption and Load-shedding Constraints*

$$\begin{cases} 0 \leq P_{wt,t} \leq P_{wt,t}^E \\ 0 \leq P_{pv,t} \leq P_{pv,t}^E \\ 0 \leq P_{cl,t} \leq P_{l,t} \end{cases} \quad (8)$$

where $P_{wt,t}^E$ and $P_{pv,t}^E$ are the expected outputs of the WT and PV.

*B. Stage 2: CIES Flexibility Margin Modeling*

Based on the pre-dispatch results, the maximum and minimum flexible supply abilities of CIES can be calculated for each dispatch state according to Eq. (9). Notably, since only electricity is traded between the ADN and CIES, only electricity subsystem flexibility supplies are considered here.

$$\begin{aligned} \max F_{CIES,t} &= F_{mt,t} + F_{bes,t} - F_{eb,t}^{in} - F_{tse,t} - F_{ec,t}^{in} \\ \min F_{CIES,t} &= F_{mt,t} + F_{bes,t} - F_{eb,t}^{in} - F_{tse,t} - F_{ec,t}^{in} \end{aligned} \quad (9)$$

where $F_{CIES,t}$, $F_{mt,t}$, $F_{bes,t}$, $F_{eb,t}^{in}$, $F_{tse,t}$, and $F_{ec,t}^{in}$ are the flexible supplies of the CIES, MT, BES, EB, TSE load, and EC.

The constraints of Stage 2 are as follows:

*1) Heat and Cooling Power Balance Constraints*

$$\begin{cases} F_{hrb,t}^{in}\eta_{hrb} + F_{eb,t}^{in}\eta_{eb} + F_{hes,t}^{dc} = H_{l,t} + F_{ac,t}^{in} + F_{hes,t}^{ch} \\ F_{ac,t}^{in}\eta_{ac} + F_{ec,t}^{in}\eta_{ec} + F_{ces,t}^{dc} = C_{l,t} + F_{ces,t}^{ch} \end{cases} \quad (10)$$

where $F_{hrb,t}^{in}$ and $F_{ac,t}^{in}$ are the flexible inputs of the HRB and AC during period $t$; $F_{hes,t}^{dc}$, $F_{hes,t}^{ch}$ and $F_{ces,t}^{dc}$, $F_{ces,t}^{ch}$ are the flexible discharge and charge powers of heat energy storage (HES) and cooling energy storage (CES).

*2) Flexible Resource Supply Constraints*

The flexibility supplies of flexible resources need to be within the adjustable range of their outputs.

$$F_{i,t}^l \leq F_{i,t} \leq F_{i,t}^u \quad (11)$$

where $F_{i,t}^l$ and $F_{i,t}^u$ are the upper and lower flexible supplies of flexible resource $i$ during period $t$, which are modeled as described below.

The flexible supply boundaries of energy-producing and coupling devices are constrained by the ramp rate and output limits as follows:

$$\begin{cases} F_{i,t}^l = \max\{P_{i,t-1}^* - r_i^d \Delta t, P_i^{\min}\} \\ F_{i,t}^u = \min\{P_i^{\max}, P_{i,t-1}^* + r_i^u \Delta t\} \end{cases} \quad (12)$$

where $P_{i,t-1}^*$ is the solution result of $P_{i,t-1}$ in Stage 1.

For energy storage devices, the flexible supply boundaries are constrained by the power and capacity limits [20]:

$$\begin{cases} F_{i,t}^l = \min\{P_i^{ch,\max}, [S_{i,t}^{\max} - (1-k_{i,loss})S_{i,t-1}^*]/\eta_i^{ch}\} \\ F_{i,t}^u = \min\{P_i^{dc,\max}, [(1-k_{i,loss})S_{i,t-1}^* - S_i^{\min}] \times \eta_i^{dc}\} \end{cases} \quad (13)$$

For TSE load, except for the ratio limits of the shifted-in and shifted-out loads, the total amount of shifted-in and shifted-out loads should be consistent within a dispatch cycle [21]. Therefore, the flexible supply boundaries can be modeled as:

$$\begin{cases} F_{tse,t}^l = \max\left\{-\alpha_{tse} P_{l,t}, \min(-\alpha_{tse} \sum_{\tau=t+1}^T P_{l,\tau} - \sum_{\tau=1}^{t-1} P_{tse,\tau}^*, 0)\right\} \\ F_{tse,t}^u = \min\left\{\alpha_{tse} P_{l,t}, \max(\alpha_{tse} \sum_{\tau=t+1}^T P_{l,\tau} - \sum_{\tau=1}^{t-1} P_{tse,\tau}^*, 0)\right\} \end{cases} \quad (14)$$

Solving the above model can obtain the flexible supply interval of CIES for each period, then the upward and downward flexibility margins for each flexible resource $i$ during period $t$ can be calculated as:

$$\begin{aligned} M_{i,t}^+ &= F_{i,t} \mid \max F_{CIES,t} - P_{i,t} \\ M_{i,t}^- &= P_{i,t} - F_{i,t} \mid \min F_{CIES,t} \end{aligned} \quad (15)$$

where $M_{i,t}^+$ and $M_{i,t}^-$ are the upward and downward flexibility margins of flexible resource $i$ during period $t$; $F_{i,t} \mid \max F_{CIES,t}$ and $F_{i,t} \mid \min F_{CIES,t}$ are the flexible outputs of flexible resource $i$ when the CIES maximizes and minimizes the flexible supply.

Subsequently, the upward and downward flexibility margins of the CIES during period $t$ are:

$$\begin{aligned} M_{CIES,t}^+ &= \sum M_{i,t}^+ \\ M_{CIES,t}^- &= \sum M_{i,t}^- \end{aligned} \quad (16)$$

## III. FAS PRICING STRATEGY

With the large-scale distributed REGs integration into ADNs, the flexibility shortage faced by ADNs has intensified. In this paper, considering that MCIESs have abundant flexible resources, the flexibility adjustment service proactively provided by MCIESs to ADNs is defined as FAS. The FAS price has a significant impact on flexibility trading, therefore determining the FAS price is crucial.

### A. CIES Adjustment Cost for FAS

FAS price is set on the basis of the economic losses incurred by CIES in response to flexibility adjustments, so the CIES adjustment cost should be determined first. Due to CIES only providing electricity regulation for ADN, the CIES adjustment cost is mainly the unit generation cost of generator set. The cost of unit electricity $P_{mt,e}^{unit}$ generated by the generators in CIES, i.e., MTs, is $\gamma_{mt}$, which includes fuel and O&M costs:

$$\gamma_{mt} = \gamma_{gas}(P_{mt,e}^{unit}/\eta_{mt}/H_{gas}) + \kappa_{mt} P_{mt,e}^{unit} \quad (17)$$

Since MT can supply both electricity and heat, it produces electricity $P_{mt,e}^{unit}$ while outputting thermal power $P_{mt,h}^{unit}$. However, the thermal power is supplied to the internal heat load of CIES through HRB and does not participate in FAS, so the adjustment cost should be obtained by subtracting the heating cost $\gamma_{mt,h}$ from $\gamma_{mt}$. For calculating $\gamma_{mt,h}$, it is assumed that all the power produced by MT is used for heating, as shown in Fig. 2. The electrical power generated by MT is converted into thermal power through EB, and the thermal power directly generated by MT is heated through HRB.

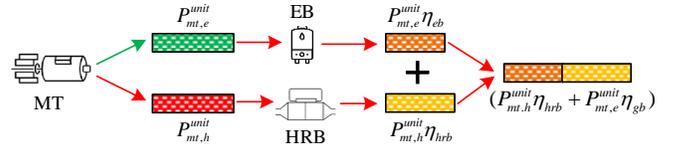

**Fig. 2.** Schematic of the MT energy supply

MT can output available thermal power $(P_{mt,h}^{unit}\eta_{hrb} + P_{mt,e}^{unit}\eta_{gb})$ through EB and HRB at a cost of $\gamma_{mt}$. So the cost of producing thermal power $P_{mt,h}^{unit}$ is:

$$\gamma_{mt,h} = \gamma_{mt} \frac{P_{mt,h}^{unit}\eta_{hrb}}{P_{mt,h}^{unit}\eta_{hrb} + P_{mt}^{unit}\eta_{gb}} \quad (18)$$

Therefore, the CIES adjustment cost is:

$$\gamma_{mt,e} = \gamma_{mt} - \gamma_{mt,h} \quad (19)$$

### B. Flexibility Auxiliary Service Price

From a risk management perspective, CIES's provision of FAS carries the risk of its own lack of flexibility. To reduce this risk, the FAS price should be set based on the flexibility margin of CIES. That is, the smaller the flexibility margin of the CIES, the greater risk of insufficient flexibility it bears, and the higher the FAS price. This reduces the probability of ADN purchasing FAS, thus reducing the risk of CIES. Conversely, when the flexibility margin of CIES is large, the FAS price should be lowered to increase its probability of providing FAS. Therefore, based on the cost-oriented pricing method and incorporating the aforementioned risk, the relationship between FAS price and CIES adjustment cost as well as flexibility margin can be approximated as a linear function shown in Fig. 3 with the expression:

$$\gamma_{flex,t} = \gamma_{flex}^{\max} - \frac{\gamma_{flex}^{\max} - \gamma_{mt,e}}{\max(\{M_{CIES}^+\})} M_{CIES,t}^+ \quad (20)$$

where $\gamma_{flex}^{\max}$ is the upper limit of FAS price; $M_{CIES}^+$ denotes the set of all $M_{CIES,t}^+$ in the dispatch cycle.

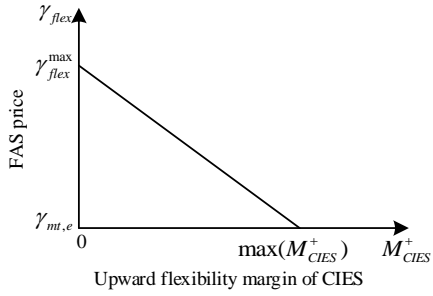

**Fig. 3.** Schematic of the FAS price

## IV. MCIES-ADN HIERARCHICAL DISPATCH MODEL

### A. Flexibility Interaction Mechanism

Due to the volatility and uncertainty of REGs and load demand, the risk of insufficient flexibility, i.e., renewable curtailment and load shedding, exists in each CIES and ADN. Scientific assessment of this risk is useful for helping decision-makers choose a rational flexibility interaction scheme. Therefore, based on the conditional value at risk (CVaR) theory in economics [22], the CVaR of insufficient flexibility is proposed and discretized by combining typical scenarios as:

$$F_{ren,t}^{CVaR} = \alpha_1 + \frac{1}{1-\beta}\sum_{s=1}^{S} p_{pro,s} P_{cr,t}$$
$$F_{load,t}^{CVaR} = \alpha_2 + \frac{1}{1-\beta}\sum_{s=1}^{S} p_{pro,s} P_{cl,t} \quad (21)$$

where $F_{ren,t}^{CVaR}$ and $F_{load,t}^{CVaR}$ are the CVaR of renewable curtailment and load shedding (i.e., the expected value that exceeds the flexibility deficit threshold at a given confidence level $\beta$); $\alpha_1$ and $\alpha_2$ are the risk thresholds of renewable curtailment and load shedding; $S$ is the number of stochastic optimization scenarios; $p_{pro,s}$ is the probability of scenario $s$.

Based on the CVaR of insufficient flexibility, the following MCIES-ADN flexibility interaction mechanism is designed:

1) Flexible resources are supplied by CIES to AND at FAS price when the CIES flexibility is sufficient and the ADN flexibility is insufficient;

2) Flexible resources are provided by ADN to CIES at time-of-use (TOU) price when the ADN flexibility is sufficient and the CIES flexibility is insufficient;

3) The CIES and ADN perform flexibility interaction based on economics when both flexibilities are sufficient;

4) The CIES and ADN perform flexibility interaction based on the CVaR of insufficient flexibility when both flexibilities are insufficient.

### B. CIES Dispatch Model

*1) Objective Function*

The objective function of CIES optimal dispatch adds the flexibility interaction cost with the ADN to the pre-dispatch cost. The subscript $n$ for the CIES is omitted because of the consistent scheduling model for each CIES.

$$\min C_{CIES} = C_{CIES}^{pre} + \sum_{s=1}^{S} p_{pro,s} \sum_{t=1}^{T} (\gamma_{TOU,t} P_{ADN,s,t}^{buy} - \gamma_{flex,t} P_{ADN,s,t}^{sell}) \Delta t \quad (22)$$

where $\gamma_{TOU,t}$ is the TOU price; $P_{ADN,s,t}^{buy}$ and $P_{ADN,s,t}^{sell}$ are the flexible resources purchased and sold from ADN under scenario $s$, respectively.

*2) Constraints*

The constraints of (2)-(8) in the CIES pre-dispatch model should also be satisfied here. However, the balance constraint of electric power in (7) needs to be modified as follows:

$$P_{wt,s,t} + P_{pv,s,t} + P_{mt,s,t} + P_{es,s,t}^{dc} + P_{ADN,s,t}^{buy}$$
$$= P_{l,s,t} + P_{tse,s,t} + P_{es,s,t}^{ch} + P_{eb,s,t}^{in} + P_{ec,s,t}^{in} - P_{cl,s,t} + P_{ADN,s,t}^{sell} \quad (23)$$

In addition, the flexibility interaction between the CIES and ADN, as well as the CVaR of insufficient flexibility, should satisfy the following constraints:

*a) Flexible interaction constraints*

$$0 \leq P_{ADN,s,t}^{buy} \leq \delta_{ADN,s,t}^{buy} P^{buy,\max}$$
$$0 \leq P_{ADN,s,t}^{sell} \leq \delta_{ADN,s,t}^{sell} P^{sell,\max} \quad (24)$$
$$\delta_{ADN,s,t}^{buy} + \delta_{ADN,s,t}^{sell} \leq 1$$

where $\delta_{ADN,s,t}^{buy}$ and $\delta_{ADN,s,t}^{sell}$ are the 0-1 variables representing the power purchase and sale status of the CIES, respectively.

*b) CVaR of insufficient flexibility constraints*

$$F_{CIES,ren,s,t}^{CVaR} \leq \phi_{CIES,s,t}^{u}$$
$$F_{CIES,load,s,t}^{CVaR} \leq \phi_{CIES,s,t}^{d} \quad (25)$$

where $\phi_{CIES,s,t}^{u}$ and $\phi_{CIES,s,t}^{d}$ are the permissible upper limits of the CVaR for the upward and downward insufficient flexibilities of CIES during period $t$ under scenario $s$.

### C. ADN Dispatch Model

*1) Objective Function*

The minimum operating cost is adopted as the objective function of ADN to be optimized, which is expressed as follows:

$$\min C_{ADN} = C_{ADN}^1 + C_{ADN}^2 + C_{ADN}^3 + C_{ADN}^4 + C_{ADN}^5$$

$$C_{ADN}^1 = \sum_{s=1}^{S} p_{pro,s} \sum_{t=1}^{T} \gamma_{GAS} (P_{MT,s,t} / \eta_{MT} / H_{gas}) \Delta t$$

$$C_{ADN}^2 = \sum_{s=1}^{S} p_{pro,s} \sum_{t=1}^{T} \sum_{i \in \Lambda_D} P_{i,s,t} \kappa_i \Delta t \quad (26)$$

$$C_{ADN}^3 = \sum_{s=1}^{S} p_{pro,s} \sum_{t=1}^{T} \sum_{n=1}^{N} (\gamma_{flex,t} P_{CIES,n,s,t}^{buy} - \gamma_{TOU,t} P_{CIES,n,s,t}^{sell}) \Delta t$$

$$C_{ADN}^4 = \sum_{s=1}^{S} p_{pro,s} \sum_{t=1}^{T} \gamma_{main,t} P_{main,s,t} \Delta t$$

$$C_{ADN}^5 = \sum_{s=1}^{S} p_{pro,s} \sum_{t=1}^{T} (\gamma_{CR} P_{CR,s,t} + \gamma_{CL} P_{CL,s,t}) \Delta t$$

where $C_{ADN}^1$, $C_{ADN}^2$, $C_{ADN}^3$, $C_{ADN}^4$, $C_{ADN}^5$ represent the fuel cost of MT, O&M cost, interaction cost with the MCIES, power purchase cost from the main grid, and insufficient flexibility penalty cost; $P_{MT,s,t}$ and $\eta_{MT}$ are the output and efficiency of MT in the ADN during period $t$ under scenario $s$; $\Lambda_d$ is the set of distributed generators in the ADN, including WT, PV, MT, and BES, $P_{i,s,t}$ and $\kappa_i$ are their outputs and O&M cost factors; $P_{main,s,t}$ and $\gamma_{main,t}$ are the power purchased from the main grid and price, respectively; $P_{CR,s,t}$ and $P_{CL,s,t}$ are the renewable curtailment and load shedding during period $t$ under scenario $s$;

$N$ is the number of CIESs connected in the ADN.

*2) Constraints*

*a) AC power flow constraints*

$$\sum_{k\in\Omega_L(j,:)} P_{jk,s,t} - \sum_{i\in\Omega_L(:,j)} (P_{ij,s,t} - I_{ij,s,t}^2 R_{ij})$$
$$= \sum_{r\in\Omega_{DG}} P_{j,s,t}^r - (P_{L,j,s,t} - P_{CL,j,s,t}) + \sum_{n=1}^{N} (P_{CIES,n,s,t}^{buy} - P_{CIES,n,s,t}^{sell})$$
$$\sum_{k\in\Omega_L(j,:)} Q_{jk,s,t} - \sum_{i\in\Omega_L(:,j)} (Q_{ij,s,t} - I_{ij,s,t}^2 X_{ij}) = Q_{L,j,s,t} \quad (27)$$
$$U_{j,s,t}^2 = U_{i,s,t}^2 - 2(R_{ij}P_{ij,s,t} + X_{ij}Q_{ij,s,t}) + (R_{ij}^2 + X_{ij}^2)I_{ij,s,t}^2$$
$$I_{ij,s,t}^2 = (P_{ij,s,t}^2 + Q_{ij,s,t}^2)/U_{i,s,t}^2$$

where $\Omega_L(j,:)$ is the set of end nodes of the branches with $j$ as the first node and $\Omega_L(:,j)$ is the set of first nodes of the branches with $j$ as the end node; $\Omega_{DG}$ is the set of distributed generators; $P_{jk,s,t}$, $P_{ij,s,t}$ and $Q_{jk,s,t}$, $Q_{ij,s,t}$ are the active and reactive powers of lines $jk$ and $ij$ during period $t$ under scenario $s$, respectively; $X_{ij}$ is the reactance of line $ij$; $P_{j,s,t}^r$ is the active power injected into node $j$ during period $t$ by distributed generator $r$; $P_{L,j,s,t}$ and $Q_{L,j,s,t}$ are the original active and reactive loads at node $j$ during period $t$ under scenario $s$, respectively; $U_{j,s,t}$ is the voltage magnitude at node $j$.

*b) Voltage and current constraints*

$$U_{i,\min}^2 \leq U_{i,s,t}^2 \leq U_{i,\max}^2$$
$$0 \leq I_{ij,s,t}^2 \leq I_{ij,\max}^2 \quad (28)$$

where $U_{i,\min}$ and $U_{i,\max}$ are the upper and lower voltage limits of node $i$; $I_{ij,\max}$ is the security current of line $ij$.

*c) Interaction constraint between the ADN and main grid*

$$0 \leq P_{main,s,t} \leq P_{main}^{\max} \quad (29)$$

where $P_{mian}^{\max}$ is the maximum power purchased from main grid.

In addition to the above constraints, the power output constraints of distributed generators, renewable curtailment and load shedding constraints, flexibility interaction constraints with MCIESs, and CVaR of insufficient flexibility constraints, are similar to those in (2), (5), (8), (24), and (25), and thus they are not repeated here.

## V. MODEL SOLVING

### A. Second-order Cone Relaxation Transformations

The AC power flow constraints in ADN with nonconvex and nonlinear structures are subjected to second-order cone relaxation so that the original ADN model can be transformed into second-order cone programming form, which can be solved efficiently using analytical methods. The details can be found in our previous work [23].

### B. ATC-based Distributed Solving

ATC is a mathematical method proposed in recent years for the optimization of hierarchical systems with superior convergence. The basic concept is that the systems are coupled with each other through decision variables [24]. In the proposed model, the ADN and MCIES are coupled through the tie-line powers. After the initial optimal dispatch of the ADN, the optimized tie-line power, $\overline{P}_{ADN,n,t}$, is passed on as a known parameter to the MCIES. Subsequently, a penalty function is added to the objective function of each CIES to reduce the deviation of the interaction power from $\overline{P}_{ADN,n,t}$ as follows:

$$\min C_{CIES_n} + \sum_{t=1}^{T} \omega_{n,t} \left| P_{ADN,n,t}^{buy} - P_{ADN,n,t}^{sell} - \overline{P}_{ADN,n,t} \right| \quad (30)$$

where $\omega_{n,t}$ is the penalty function multiplier of CIES $n$.

Similarly, after the optimization of each CIES is completed, the tie-line power, $\overline{P}_{CIES,n,t}$, is transmitted to the ADN. A penalty function is also added to the objective function of ADN to reduce the deviation of its interaction power from $\overline{P}_{CIES,n,t}$:

$$\min C_{ADN} + \sum_{t=1}^{T}\sum_{n=1}^{N} \omega_{n,t} \left| P_{CIES,n,t}^{buy} - P_{CIES,n,t}^{sell} - \overline{P}_{CIES,n,t} \right| \quad (31)$$

This process is repeated until the following convergence criteria are satisfied:

$$\left| P_{ADN,n,t}^{buy,k} - P_{ADN,n,t}^{sell,k} - (P_{CIES,n,t}^{buy,k} - P_{CIES,n,t}^{sell,k}) \right| \leq \varepsilon_1$$
$$\left| \frac{(C_{ADN}^k + \sum_{n=1}^{N} C_{CIES_n}^k) - (C_{ADN}^{k-1} + \sum_{n=1}^{N} C_{CIES_n}^{k-1})}{(C_{ADN}^{k-1} + \sum_{n=1}^{N} C_{CIES_n}^{k-1})} \right| \leq \varepsilon_2 \quad (32)$$

Equation (32) indicates that in the $k$th iteration, the difference between the tie-line powers of the MCIES and the ADN should satisfy the accuracy requirement. In addition, the overall benefits of the ADN containing MCIESs are checked for optimization.

If the convergence criteria cannot be satisfied simultaneously, the penalty function multipliers are updated according to (33).

$$\omega_{n,t}^k = \omega_{n,t}^{k-1} + \chi_{n,t}^{k-1} \left| P_{ADN,n,t}^{buy,k} - P_{ADN,n,t}^{sell,k} - (P_{CIES,n,t}^{buy,k} - P_{CIES,n,t}^{sell,k}) \right|$$
$$\chi_{n,t}^k = \lambda \chi_{n,t}^{k-1} \quad (33)$$

where $\chi_{n,t}^k$ is the first-order term multiplier of the $k$th iteration during period $t$; $\lambda$ is the iterative multiplier, which is generally in the range of $2 < \lambda < 3$.

## VI. CASE STUDY

A PG&E 69-node distribution system is used for simulation analyses to validate the effectiveness of the proposed model, as shown in Fig. 4. Two WT farms with a capacity of 800 kW are located at Nodes 38 and 41, and two PV farms with a capacity of 600 kW are located at Nodes 60 and 61. The forecasted outputs of the REG are the same for all of the network nodes [25]. Nodes 17 and 40 are equipped with an MT and a BES, respectively. A residential CIES, commercial CIES, and industrial CIES are connected to Nodes 22, 31, and 44, respectively. The ADN parameters are listed in Table I, the MCIES parameters are listed in Table II and [26], the TOU and main grid prices are shown in Fig. 5. The number of stochastic optimization scenarios is obtained as five by Monte Carlo simulation and K-means clustering. The initial values of $\chi_n$ and $\omega_n$ are 0.0001 and 0.02, $\varepsilon_1$ and $\varepsilon_2$ are set to 0.01, $T=24$ h, and $\Delta t=1$ h. All procedures are performed using MATLAB R2018a with the CPLEX solver.

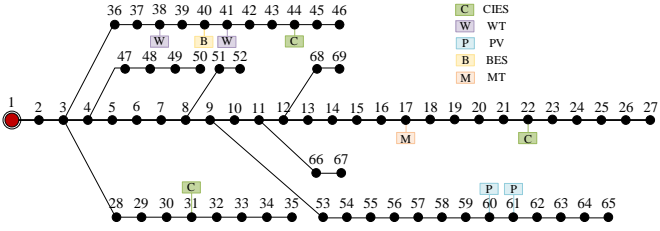

**Fig. 4.** PG&E 69-node system diagram

TABLE I
ADN PARAMETERS

| Parameter | Value | Parameter | Value |
|---|---|---|---|
| $P_{MT}^{max}$ (kW) | 2500 | $\gamma_{GAS}$ ($/m$^3$) | 0.4 |
| $P_{MT}^{min}$ (kW) | 200 | $H_{gas}$ (kWh/m$^3$) | 9.7 |
| $r_{MT}^{d}/r_{MT}^{u}$ (kW) | 1200 | $\gamma_{CR}$ ($/kW) | 0.1 |
| $T_{MT}^{up}$ (h) | 4 | $\gamma_{CL}$ ($/kW) | 0.2 |
| $P_{BES}^{max}$ (kW) | 350 | $\phi_{ADN}^{d}/\phi_{ADN}^{u}$ (kW) | 500 |
| $S_{BES}^{min}$ (kWh) | 300 | $\eta_{BES}^{ch}$ | 0.96 |
| $S_{BES}^{max}$ (kWh) | 1500 | $\eta_{BES}^{dc}$ | 0.96 |
| $S_{BES}^{0}$ (kWh) | 750 | $\beta_{ADN}$ | 0.95 |

TABLE II
MCIES PARAMETERS

| Parameter | Residential CIES | Commercial CIES | Industrial CIES |
|---|---|---|---|
| $P_{mt}^{max}$ (kW) | 1000 | 650 | 1400 |
| $P_{mt}^{min}$ (kW) | 200 | 100 | 200 |
| $r_{mt}^{d}/r_{mt}^{u}$ (kW) | 500 | 300 | 700 |
| $P_{eb}^{max}$ (kW) | 400 | 400 | 900 |
| $r_{eb}^{d}/r_{eb}^{u}$ (kW) | 200 | 200 | 400 |
| $P_{ac}^{max}$ (kW) | 500 | 500 | 800 |
| $P_{ac}^{min}$ (kW) | 60 | 60 | 80 |
| $r_{ac}^{d}/r_{ac}^{u}$ (kW) | 200 | 200 | 400 |
| $P_{ec}^{max}$ (kW) | 100 | 200 | 500 |
| $r_{ec}^{d}/r_{ec}^{u}$ (kW) | 50 | 100 | 200 |
| $\phi_{CIES}^{u}/\phi_{CIES}^{d}$ (kW) | 200 | 200 | 500 |
| $\gamma_{gas}$ ($/m$^3$) | 0.45 | 0.55 | 0.55 |
| $\beta_{CIES}$ | 0.95 | 0.95 | 0.95 |

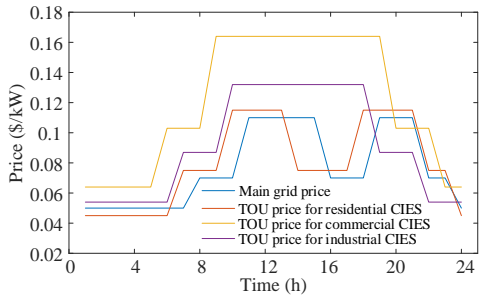

**Fig. 5.** TOU price and main grid price

### A. CIES Flexibility Evaluation Analysis

To verify the effectiveness of the CIES flexibility evaluation, the residential CIES is considered as an example, and the upward and downward flexibility margins for pre-dispatch are evaluated, as shown in Fig. 6.

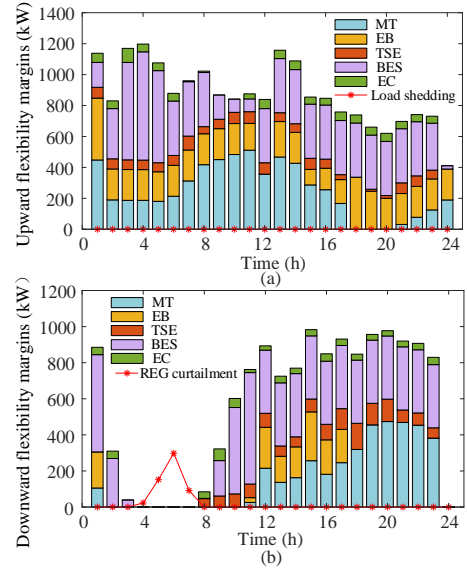

**Fig. 6.** Upward (a) and downward (b) flexibility margins for pre-dispatch of the residential CIES

As shown in Fig. 6, the upward flexibility margins of the residential CIES are sufficient in the pre-dispatch phase, and no load shedding occurs during the dispatch cycle. The downward flexibility margins are zero from 4:00–7:00, and REG curtailment occurs during this period. In addition, the downward flexibility margin is also zero at 24:00; however, no wind curtailment occurs. This is because the system is in a critical state of sufficient and insufficient flexibility at this moment due to multiple operation and energy coupling constraints, in particular the capacity conservation constraints of energy storage devices and TSE loads. This illustrates that the flexibility evaluation model can effectively quantify the flexibility adjustment capabilities of diversified flexible resources and accurately evaluate the flexibility margins of CIESs with complex energy coupling.

### B. FAC Price Analysis

After evaluating the flexibility margins of MCIES, the FAS prices for each CIES are obtained according to the FAS pricing strategy, which are shown in Fig. 7. It can be seen the smaller the upward flexibility margin of CIES, the higher the FAS price, and vice versa. This indicates that the proposed FAS pricing strategy effectively combines the flexibility of MCIESs, so that MCIESs are able to sell flexible resources rationally according to their flexibility margins. In addition, all of the FAS prices are higher than the adjustment cost, which ensures the profitability of MCIES.

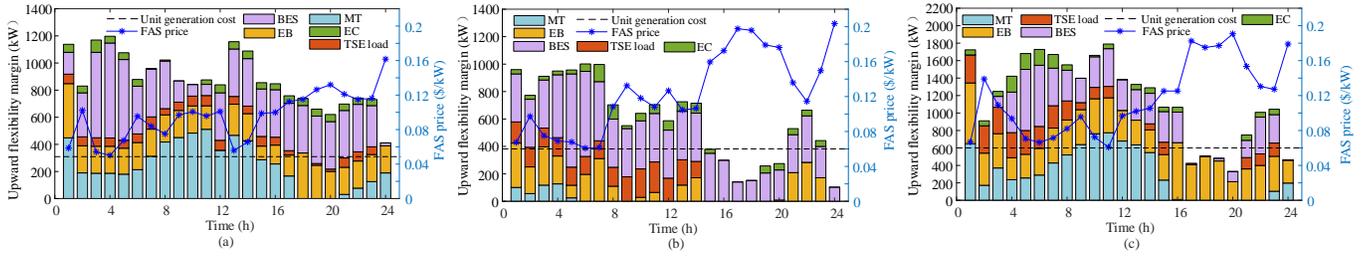

**Fig. 7.** FAS prices of residential CIES (a), commercial CIES (b), and industrial CIES (c)

*C. Dispatch Scheme Analysis*

The dispatch schemes are demonstrated using an example of the ADN and commercial CIES, as shown in Figs. 8 and 9.

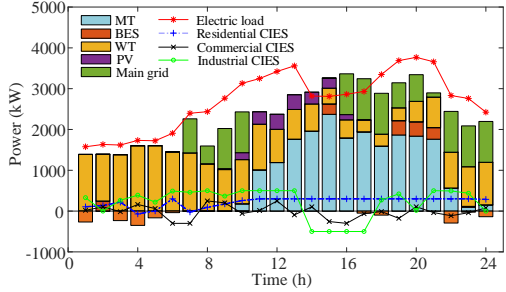

**Fig. 8.** Dispatch schemes of the ADN

As shown in Fig. 8, the ADN purchases flexible power from MCIES during the peak WT output and valley electric load period of 1:00–6:00, which avoids the start-up of MT for energy supply and reduces the electricity purchased from the main grid. The residential CIES sells electricity to the ADN during most periods owing to generally high upward flexibility margins, and the commercial CIES purchases electricity from the ADN during the period of 15:00–19:00 when the upward flexibility margins are extremely low. In addition, although the upward flexibility margins of the industrial CIES are relatively low from 16:00 to 20:00, the electricity purchased from 14:00 to 17:00 enables it to sell electricity during periods of high FAS prices to increase profits. Simultaneously, the outputs of MT in ADN are reduced owing to the electricity purchased. This indicates that the FAS improves the operational flexibility of the ADN and reduces the dependence on the main grid.

Fig. 9 shows that the MT is the main generator of electricity and heat load supply, and the EC is the main device of the cooling load supply owing to its high-performance coefficient. In addition, during the period of valley electricity load and peak heat load from 1:00 to 5:00, the WT outputs are fully absorbed and supplied to the EB for heating, and FAS is provided to the ADN to improve revenue. During the period of valley REG outputs and peak electricity loads from 15:00 to 19:00, the commercial CIES purchases electricity from the ADN to meet the electricity demand. In summary, the CIES utilizes surplus flexible resources to supply the ADN and gain revenue, which realizes a double win for the MCIES and ADN.

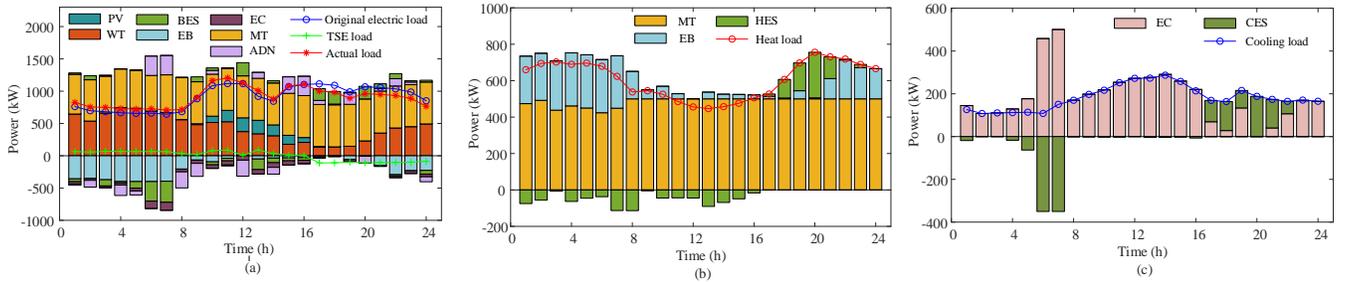

**Fig. 9.** Dispatch schemes of the commercial CIES

*D. Mode Comparative Analysis*

To verify the validity and superiority of the proposed model, the following four modes are considered for comparison.

Mode 1: ADN-MCIES optimal dispatch without considering the FAS pricing strategy; instead, the FAS is conducted using the TOU price.

Mode 2: ADN-MCIES optimal dispatch without considering the risk of insufficient flexibility.

Mode 3: ADN-MCIES optimal dispatch considering the FAS pricing strategy and the risk of insufficient flexibility using the centralized modeling and solution approach.

Mode 4: ADN-MCIES optimal dispatch considering the FAS pricing strategy and the risk of insufficient flexibility using the distributed modeling and solution approach.

*1) FAS Pricing Strategy Analysis:*

The operating costs and amount of insufficient flexibility of the system under Modes 1 and 4 are presented in Table III.

TABLE III
OPERATING COSTS AND AMOUNT OF INSUFFICIENT FLEXIBILITY

| Mode | Operating cost of ADN ($) | Operating cost of MCIES ($) | Total cost ($) | Amount of insufficient flexibility (kW) |
|---|---|---|---|---|
| Mode 1 | 3551.20 | 9687.91 | 13239.11 | 114.34 |
| Mode 4 | 3814.43 | 9189.82 | 13004.25 | 0 |

The results in Table III indicate that compared to Mode 1, the operating cost of MCIES decreases and the cost of ADN increases under Mode 4; however, the total cost decreases, and the risk of insufficient flexibility is avoided. This is because the MCIES has plentiful flexible resources to enhance the system operational flexibility. Therefore, surplus flexible resources are available to supply the ADN, which increases the revenue from electricity sales. Besides, the FAS pricing strategy in Mode 4 combines the flexibility margins of the MCIES and thus avoids the risk of insufficient flexibility.

Additionally, the flexibility interactions between the MCIES and ADN under Modes 1 and 4 are compared using the industrial CIES as an example, as shown in Fig. 10.

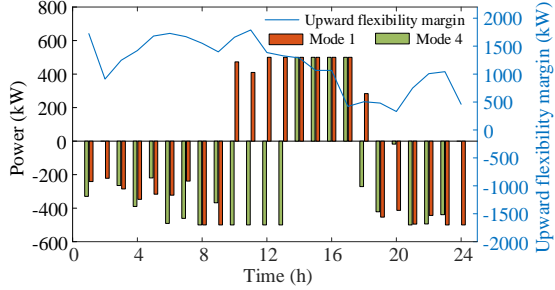

**Fig. 10.** Flexibility interaction between the industrial CIES and ADN

Fig. 10 shows that during the period of 10:00–13:00 when the upward flexibility margins of the industrial CIES are sufficient, the CIES in Mode 4 provides flexible resources to the ADN, while purchases flexible resources from the ADN in Mode 1. Besides, the flexible resources sold to the ADN in Mode 4 are significantly reduced at 20:00 when the upward flexibility margin is the lowest during the dispatch cycle. This shows that the proposed FAS pricing strategy can effectively incorporate the flexibility margins of CIES to rationalize the FAS transactions between the ADN and MCIES.

*2) Analysis of the Risk of Insufficient Flexibility*

Because the flexibility interaction of the three CIESs with the ADN is considered, the flexibility of the overall ADN is sufficient and there is no risk of insufficient flexibility. To verify the effectiveness of the proposed ADN-CIES interaction strategy based on the risk of insufficient flexibility, only the commercial CIES is considered to access the ADN, and the CVaR of insufficient flexibility under Modes 2 and 4 are compared, as shown in Fig. 11.

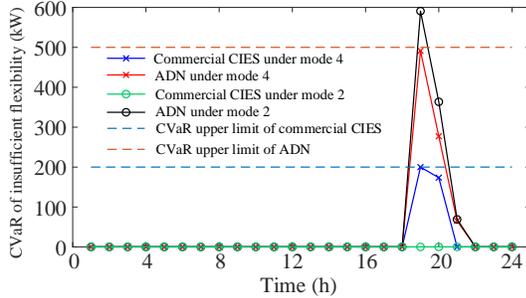

**Fig. 11.** CVaR of insufficient flexibility under Modes 2 and 4

As shown in Fig. 11, in Mode 2, which does not consider the risk of insufficient flexibility, the ADN is sacrificed to take on more risk of insufficient flexibility to ensure operational flexibility of the commercial CIES. In contrast, the ADN and commercial CIES in Mode 4 share the risk under the ADN-CIES interaction strategy. Although the total amount of risks has increased, the risks of both are within their allowable range at each time period.

*3) Modeling Method Analysis*

A comparison between centralized modeling in Mode 3 and ATC-based distributed modeling in Mode 4 is shown in Table IV.

TABLE IV
COSTS AND CALCULATION TIMES OF DIFFERENT MODELING APPROACHES

| Mode | Operating cost of ADN ($) | Operating cost of MCIES ($) | Total cost ($) | Calculation time (s) |
|---|---|---|---|---|
| Mode 3 | 4031.31 | 8601.40 | 12632.71 | 647.31 |
| Mode 4 | 3814.43 | 9189.82 | 13004.25 | 559.49 |

Table IV reveals that the operating cost of ADN under the centralized modeling approach is relatively high, while that of MCIES is lower, resulting in a lower total operating cost than that of distributed modeling. Although the distributed modeling approach does not achieve the optimal overall operating cost of the ADN and MCIES, it refines the interest game process of multiple stakeholders, which is more consistent with the actual situation, and realizes decoupling and parallel solving for complex models. Moreover, because of the small number of CIESs in this study, the distributed modeling approach requires slightly less time than the centralized approach; however, as the number of incorporated CIESs increases, the difference in computation time between the two methods will widen further.

*E. ATC Algorithm Analysis*

The ATC-based model solution in this study reaches convergence after 10 iterations. Fig. 12 shows the variation in the maximum gap between the ADN and MCIES for each iteration. It can be seen that the gap between the two continuously converges to zero during the iteration process, which illustrates the effectiveness of the ATC algorithm.

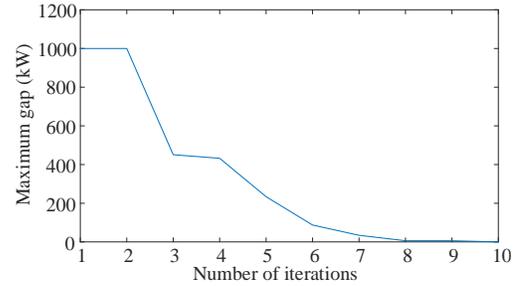

**Fig. 12.** ATC iteration curve

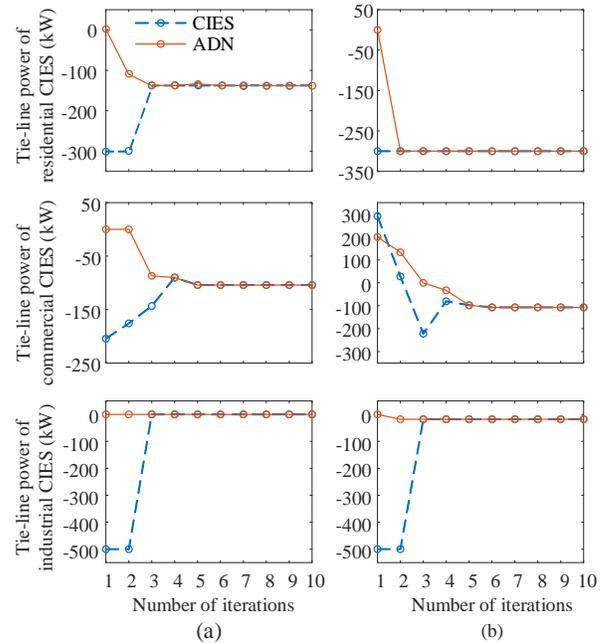

**Fig. 13.** Iterative trends of the tie-line powers at 2:00 (a) and 20:00 (b)

Fig. 13 shows the iterative convergence process of the tie-line powers between the ADN and each CIES at 2:00 and 20:00. As shown in Fig. 13, the values of the connection variables of the ADN and each CIES remain close to each other during the iterative solution process until the convergence condition is satisfied.

## VII. Conclusion

With the increasing coupling of different energy sources, various types of CIESs are gradually being incorporated into ADNs. Aiming to capture the characteristics of MCIES with plentiful flexible and controllable resources, this study presents a hierarchical optimal dispatch model for ADN considering the flexibility auxiliary services of MCIESs. The following conclusions are drawn based on the simulation results.
1) The proposed hierarchical optimal dispatch model effectively exploits the flexibility regulation potential of MCIES for ADN by formulating a FAS pricing strategy that incorporates flexibility margins, which manages to improve the operational flexibility and economy of the overall ADN containing MCIES.
2) The developed MCIES-ADN flexibility interaction mechanism reduces the uncertainty risk of REG outputs through the risk constraint of insufficient flexibility and allocates risks when both are insufficiently flexible to achieve reasonable interaction among multiple stakeholders.
3) The ATC-based distributed modeling method realizes the decoupling and parallel solving of all stakeholders within an effective time, thus truly reflecting the advantages of the decentralized autonomy of MCIES and protecting the privacies of all parties; Also, the proposed algorithm is proved to have good convergence.

Future work will focus on the impact of the MCIES-ADN flexibility interaction strategy on ADN planning. In addition, this paper has not considered the variability in the REG outputs at different nodes and interactions among MCIESs, while a more realistic scenario shall be taken into account during optimization.